\documentclass[%twocolumn,
superscriptaddress,aps,preprintnumbers,amsmath,showpacs,amssymb,prd,nofootinbib,preprint]{revtex4-1}

\usepackage[dvipdfm]{graphicx}
\usepackage{amsfonts}
\usepackage[mathscr]{eucal}
\usepackage{ascmac}
\usepackage{epstopdf}
\usepackage{dcolumn}% Align table columns on decimal point
\usepackage{bm}% bold math
\usepackage{hyperref}
\usepackage{color}

\begin{document}

%%%%%%%%%%%%%%%%%%%%%%%%%%%%%%%%%%%%%%%%%%%

\def\a{\alpha}
\def\b{\beta}
\def\c{\varepsilon}
\def\d{\delta}
\def\e{\epsilon}
\def\f{\phi}
\def\g{\gamma}
\def\h{\theta}
\def\k{\kappa}
\def\l{\lambda}
\def\m{\mu}
\def\n{\nu}
\def\p{\psi}
\def\q{\partial}
\def\r{\rho}
\def\s{\sigma}
\def\t{\tau}
\def\u{\upsilon}
\def\v{\varphi}
\def\w{\omega}
\def\x{\xi}
\def\y{\eta}
\def\z{\zeta}
\def\D{\Delta}
\def\G{\Gamma}
\def\H{\Theta}
\def\L{\Lambda}
\def\F{\Phi}
\def\P{\Psi}
\def\S{\Sigma}

\def\o{\over}
\def\beq{\begin{eqnarray}}
\def\eeq{\end{eqnarray}}
\newcommand{\gsim}{ \mathop{}_{\textstyle \sim}^{\textstyle >} }
\newcommand{\lsim}{ \mathop{}_{\textstyle \sim}^{\textstyle <} }
\newcommand{\vev}[1]{ \left\langle {#1} \right\rangle }
\newcommand{\bra}[1]{ \langle {#1} | }
\newcommand{\ket}[1]{ | {#1} \rangle }
\newcommand{\EV}{ {\rm eV} }
\newcommand{\KEV}{ {\rm keV} }
\newcommand{\MEV}{ {\rm MeV} }
\newcommand{\GEV}{ {\rm GeV} }
\newcommand{\TEV}{ {\rm TeV} }
\newcommand{\1}{\mbox{1}\hspace{-0.25em}\mbox{l}}
\def\diag{\mathop{\rm diag}\nolimits}
\def\Spin{\mathop{\rm Spin}}
\def\SO{\mathop{\rm SO}}
\def\O{\mathop{\rm O}}
\def\SU{\mathop{\rm SU}}
\def\U{\mathop{\rm U}}
\def\Sp{\mathop{\rm Sp}}
\def\SL{\mathop{\rm SL}}
\def\tr{\mathop{\rm tr}}

\def\IJMP{Int.~J.~Mod.~Phys. }
\def\MPL{Mod.~Phys.~Lett. }
\def\NP{Nucl.~Phys. }
\def\PL{Phys.~Lett. }
\def\PR{Phys.~Rev. }
\def\PRL{Phys.~Rev.~Lett. }
\def\PTP{Prog.~Theor.~Phys. }
\def\ZP{Z.~Phys. }

\def\dd{\mathrm{d}}
\def\ff{\mathrm{f}}
\def\BH{{\rm BH}}
\def\inf{{\rm inf}}
\def\ev{{\rm evap}}
\def\eq{{\rm eq}}
\def\SM{{\rm sm}}
\def\Mpl{M_{\rm Pl}}
\def\GeV{{\rm GeV}}
\newcommand{\Red}[1]{\textcolor{red}{#1}}

\def\mDM{m_{\rm DM}}
\def\mphi{m_{\phi}}
\def\TeV{{\rm TeV}}
\def\Gphi{\Gamma_\phi}
\def\TR{T_{\rm RH}}
\def\Br{{\rm Br}}
\def\DM{{\rm DM}}
\def\Eth{E_{\rm th}}
\newcommand{\lmk}{\left(}  
\newcommand{\rmk}{\right)}
\newcommand{\lkk}{\left[}  
\newcommand{\rkk}{\right]}
\newcommand{\lhk}{\left \{ }  
\newcommand{\rhk}{\right \} }
\newcommand{\del}{\partial}  
\newcommand{\la}{\left\langle} 
\newcommand{\ra}{\right\rangle}

\newcommand{\qel}{\hat{q}_{el}}
\newcommand{\ksplit}{k_{\text{split}}}
\def\GDM{\Gamma_{\text{DM}}}
\newcommand{\half}{\frac{1}{2}}
\def\Gsplit{\Gamma_{\text{split}}}

%%%%%%%%%%%%%%%%%%%%%%%%%%%%%%%%%%%%%%%%%%%%%%%%%%%%%%%%%%%%%%%

\title{
Dark Matter Production in Late Time Reheating
}

\author{Keisuke Harigaya}
\affiliation{Kavli IPMU (WPI), TODIAS, University of Tokyo, Kashiwa, 277-8583, Japan}
\author{Masahiro Kawasaki}
\affiliation{ICRR, University of Tokyo, Kashiwa, 277-8582, Japan}
\affiliation{Kavli IPMU (WPI), TODIAS, University of Tokyo, Kashiwa, 277-8583, Japan}
\author{Kyohei Mukaida}
\affiliation{Department of Physics, Faculty of Science,
University of Tokyo, Bunkyo-ku, 133-0033, Japan}
\author{Masaki Yamada}
\affiliation{Kavli IPMU (WPI), TODIAS, University of Tokyo, Kashiwa, 277-8583, Japan}
\affiliation{ICRR, University of Tokyo, Kashiwa, 277-8582, Japan}
\begin{abstract}
We estimate dark matter density for the Universe with a reheating temperature smaller than the mass of dark matter, assuming dark matter to be a weakly interacting massive particle. During the reheating process, an inflaton decays and releases high-energy particles, which are scattered inelastically by the thermal plasma and emit many particles. Dark matters are produced through these inelastic scattering processes and pair creation processes by high-energy particles. We properly take account of the Landau-Pomeranchuk-Migdal effect on inelastic processes and show that the resultant energy density of dark matter is much larger than that estimated in the literature and can be consistent with that observed when the mass of dark matter is larger than $O(100)$ GeV.
\end{abstract}

\date{\today}
\maketitle
\preprint{IPMU 14-0027}
\preprint{ICRR-report-671-2013-20}
\preprint{UT-14-04}
%%%%%%%%%%%%%%%%%%%%%%%%%%%%%%%%%%%%%%%%%%%%%%%%%%%%%%%%%%%%%%%

%%%%%%%%%%%%%%%%%%%%%%%%%%%%%%%%%%%%%%%%%%%%%%%%%%%%%%%%%%%%%%%%
\section{\label{sec1}Introduction}
%%%%%%%%%%%%%%%%%%%%%%%%%%%%%%%%%%%%%%%%%%%%%%%%%%%%%%%%%%%%%%%%

A weakly interacting massive particle (WIMP) is one of the most attractive candidates for dark matter (DM),
motivated by new physics at a $\TeV$ scale, including supersymmetric (SUSY) theories.
DM is produced thermally in the early Universe, and its abundance can be consistent with that observed,
if the reheating temperature of the Universe is sufficiently larger than its freeze-out temperature.
The coincidence of the observed dark matter density and the relic density determined
by the weak interaction scale is referred to as the WIMP miracle.
Since this scenario requires the mass of the WIMP at the electroweak or $\TeV$ scale,
there are rich implications for near-future experiments,
including direct and indirect detection experiments of DM and particle collider experiments.

However,
when we look at each model motivated by particle physics,
it is nontrivial to obtain a correct mass spectrum that can account for the abundance of DM.
For example, in the constrained minimal SUSY standard model,
the lightest SUSY particle (LSP) is bino-like.
Non-observation of SUSY particles and the discovery of the 126~GeV Higgs boson~\cite{Aad:2012tfa,Chatrchyan:2012ufa}
by the LHC experiment indicate that SUSY particles are heavy,
which leads to overproduction of the bino-like LSP in the early Universe.%
\footnote{
This problem can be avoided by co-annihilation~\cite{Griest:1990kh,Gondolo:1990dk} with the stau~\cite{Ellis:1998kh},
but only with fine-tuning.
For recent discussion on SUSY models with a correct DM abundance,
see Refs.~\cite{PGM,Feng:2011aa,Okada:2012gf,Feng:2012rn,ArkaniHamed:2012gw,Yanagida:2013ah}, for example.
}
Taking this situation seriously,
we reconsider the assumption of high reheating temperature.

When we consider inflationary models,
a scenario with a low reheating temperature is naturally realized as follows.
The inflaton is required to have a very flat potential, which suggests some symmetry to control its potential.
The symmetry naturally suppresses interactions of the inflaton,
which in turn leads to a low reheating temperature of the Universe.
For example, if the mass of the inflaton is of the order of $10^{11}$ GeV
and it decays through a dimension six  Planck-suppressed operator,
the reheating temperature of the Universe is less than about $O(1)$ GeV,
which is smaller than typical freeze-out temperatures of WIMPs.

In a scenario with a low reheating temperature,
the thermal abundance of DM is much less than a scenario with high reheating temperature,
mainly because the energy density of the thermal plasma is
a subdominant component of that of the Universe at the time of DM freeze out~\cite{Chung:1998rq, Giudice:2000ex}.
In other words, the thermal abundance of DM is diluted by the entropy production from the inflaton decay.

However, the entropy production itself provides
DM~\cite{Moroi:1994rs, Kawasaki:1995cy, Moroi:1999zb, Allahverdi:2002nb, Gelmini:2006pw, Kurata:2012nf}.
In Ref.~\cite{Kurata:2012nf},
they have indicated that DM is produced in a shower from the decay of the inflaton
and have calculated the resultant DM energy density
using generalized Dokshitzer-Gribov-Lipatov-Altarelli-Parisi (DGLAP) equations~\cite{Sarkar:2001se, Barbot:2002ep}
in a certain SUSY model.
They have found that DM is produced efficiently through this process
when the inflaton decays into particles carrying a non-zero ${\rm SU}(3)_{\rm c}$ charge.
The DM abundance also depends on the mass of inflaton,
and the number of DM produced per one inflaton decay is typically $O(100)$ for the inflaton mass of $O(10^{12})$ GeV.
In addition, in Ref.~\cite{Allahverdi:2002nb},
it has been pointed out that DM is produced by inelastic scatterings
between the thermal plasma and high-energy particles produced by the inflaton decay.

Therefore, in a scenario with low reheating temperature,
the total amount of DM is given by the sum of the following contributions:
(suppressed) thermal production, production through a cascade shower from the inflaton decay,
and production through inelastic scatterings between high-energy particles and the thermal plasma.
The last contribution is closely related to thermalization processes in the era of reheating,
which has to be investigated in detail.

When reheating temperature is low enough,
the typical momentum of particles produced by the decay of inflaton,
which is roughly given by the mass of the inflaton,
is much larger than the (would-be) temperature of background plasma.
In this case,
the thermalization process is completed by splitting processes
through which the number of high-energy particles drastically increases~\cite{Kurkela:2011ti,Harigaya:2013vwa}
(see also Refs.~\cite{Davidson:2000er, Arnold:2002zm,Jaikumar:2002iq}).
The rate of splitting processes is suppressed by the Landau-Pomeranchuk-Migdal (LPM) effect,
which is a destructive interference effect
between emission processes~\cite{Landau-Pomeranchuk, Migdal, Gyulassy:1993hr,
Arnold:2001ba, Arnold:2001ms, Arnold:2002ja, Besak:2010fb}.
The increases in the number of high-energy particles as well as the LPM effect
should be taken into account in the estimation of the DM abundance produced through inelastic processes.

In this paper,
we calculate the abundance of DM produced from inelastic scatterings
between high-energy particles and the thermal plasma
with careful consideration on the thermalization process as mentioned above. 
The resultant DM abundance is independent of the mass of the inflaton as long as the mass is sufficiently large,
and depends mainly on the mass scale of the DM sector and reheating temperature.
We find that this is the dominant contribution to the amount of DM in a scenario 
with a low reheating temperature
when the mass of the inflaton is sufficiently large.
We should emphasize that this mechanism to produce DM is highly model independent.
Even if the decay channel of the inflaton into DM is absent in particular,
DM is produced through inelastic scatterings.
In addition,
this scenario can also account for the abundance of DM with mass of $O(1)$ PeV,
which is larger than unitarity bound of a few hundred TeV~\cite{Griest:1989wd}.
Such heavy DM might account for the recent observation of high-energy cosmic-ray neutrinos
by the IceCube experiment~\cite{Aartsen:2013jdh, Feldstein:2013kka, Esmaili:2013gha, Bai:2013nga}.

This paper is organized as follows.
In the next section, we explain how high-energy particles lose their energy in the thermal plasma
taking account of the LPM effect.
We also describe a thermal history of the Universe in our setup.
In Sec.~\ref{production},
we briefly review previous works for the thermal and non-thermal production of DM
and improve the calculation of the DM abundance from the inflaton decay
taking account of the LPM effect.
Then we discuss the relation between our scenario and other topics,
such as the free-streaming velocity of DM,
the Affleck-Dine baryogenesis, and SUSY theories.
Section~\ref{conclusion} is devoted to the conclusion.

%%%%%%%%%%%%%%%%%%%%%%%%%%%%%%%%%%%%%%%%%%%%%%%%%%%%%%%%%%%%%%%%
\section{\label{thermalization}thermalization and thermal history}
%%%%%%%%%%%%%%%%%%%%%%%%%%%%%%%%%%%%%%%%%%%%%%%%%%%%%%%%%%%%%%%%

In this section,
we consider a situation in which the inflaton with a mass of $\mphi$
decays into light particles,
and the light particles yield their energy into the thermal plasma
through elastic and inelastic scatterings.
In Sec.~\ref{energy loss}, we calculate the rate of energy loss by elastic and inelastic scatterings,
taking the LPM effect into account,
and show that
inelastic scatterings are the dominant process for the energy loss.
In Sec.~\ref{history}, we explain the evolution of the thermal plasma
in the expanding universe with low reheating temperature.

%%%%%%%%%%%%%%%%%%%%%%%%%%%%%%%%%%%%%%%%%%%%%%%%%%%%%%%%%%%%%%%%
\subsection{\label{energy loss}Interactions between high-energy particles and thermal plasma}
%%%%%%%%%%%%%%%%%%%%%%%%%%%%%%%%%%%%%%%%%%%%%%%%%%%%%%%%%%%%%%%%

In this subsection, we review thermalization processes of a high-energy particle
with energy $E_i$ in the thermal plasma with a low temperature $T$ ($\ll E_i$).
The thermalization occurs through elastic and inelastic scatterings between a high-energy particle and the thermal plasma.

\begin{figure}[t]
\centering % \begin{center}/\end{center} takes some additional vertical space
\includegraphics[width=.45\textwidth, bb=126 615 276 706]{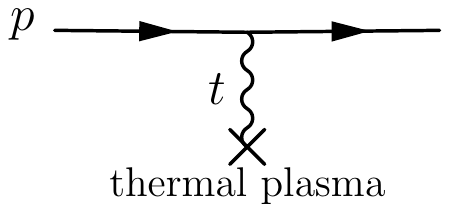} 
\caption{
Sample diagram describing an elastic scattering.
}
  \label{fig1}
\end{figure}

First, let us consider elastic scattering processes.
Figure~\ref{fig1} is one of the Feynman diagrams of elastic scattering processes.
When
the exchanged particle is a gauge boson,
the scattering cross section is dominated by the $t$-channel gauge boson exchange
and is given as
\beq
 \sigma_{\rm elastic} \sim \frac{\alpha^2}{t} \sim \frac{\alpha}{T^2},
\eeq
where $t$ is one of the Mandelstam variables and $\alpha$ is the fine-structure constant of the gauge interaction.
Although this cross section has an infrared divergence at zero temperature,
an infrared cutoff arises due to a non-zero mass of the internal gauge boson
at finite temperature and as large as $\alpha^{1/2} T$.%
\footnote{
	Strictly speaking, the total cross section still has a logarithmic divergence since the almost static magnetic
	fields are not screened perturbatively. 
	In the following discussion, we omit such logarithmic factors since they only change interaction rates weakly.
}
The rate of elastic scatterings is thus given as
\beq
 \Gamma_{\rm el} = \la \sigma_{\rm elastic} n \ra \sim \alpha T,
\eeq
where $\la \  \ra$ represents a thermal average
and $n$ ($\sim T^3$) is the number density of scattered particles in the thermal plasma.
Since the high-energy particle loses its energy by $\alpha T$ for each elastic scattering,
the energy loss rate by elastic scatterings is estimated as
\beq
 \left. \frac{\dd E}{\dd t} \right\vert_{\rm elastic} \sim \alpha T \la \sigma_{\rm elastic} n \ra \sim \alpha^2 T^2.
 \label{elastic dEdt}
\eeq
%%

%%%%%%%%%%%%%%%%%%%%%%%%%%%%%%%%%%%%%%%%%%%%%
%%%%%%%%%%%%%%%%%%%%%%%%%%%%%%%%%%%%%%%%%%%%%

\begin{figure}[t]
\centering % \begin{center}/\end{center} takes some additional vertical space
\includegraphics[width=.45\textwidth, bb=116 571 293 699]{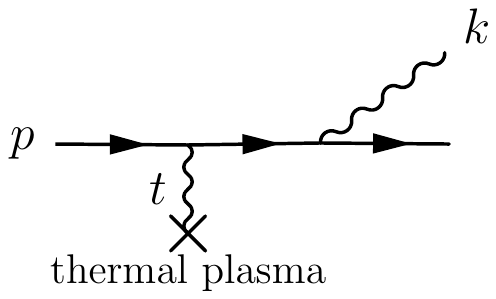} 
\caption{
Sample diagram describing an inelastic scattering.
}
  \label{fig2}
\end{figure}

Inelastic scattering cross sections are also dominated by $t$-channel contributions, 
as shown in Fig.~\ref{fig2}.
Since the intermediate fields are almost on shell (i.e., $t \sim \alpha T^2 \ll E_i^2$),
this process can be regarded as an emission
associated with an elastic scattering process.
The cross section is thus given as
\beq
 \sigma_{\rm inelastic} \sim \alpha \sigma_{\rm elastic}
 \sim \frac{\alpha^2}{T^2},
\eeq
where we implicitly assume that daughter particles are massless.
One may consider that the rate of the splitting process is simply given by $\la \sigma_{\rm inelastic} n \ra$.
However, we have to take account of an interference effect among emission processes,
known as the LPM effect~\cite{Landau-Pomeranchuk, Migdal,
Gyulassy:1993hr,
Arnold:2001ba, Arnold:2001ms, Arnold:2002ja, Besak:2010fb}.
As we show below, the rate of inelastic processes is in fact affected and suppressed by the LPM effect.

%%%%%%%%%%%%%%%%%%%%%%%%%%%%%%%%%%%%%%%%%%%%%
%%%%%%%%%%%%%%%%%%%%%%%%%%%%%%%%%%%%%%%%%%%%%

Here, we briefly explain how the interference and suppression for
inelastic scatterings occur, following Ref.~\cite{Gyulassy:1993hr}.
Let us consider classical electrodynamics as an illustration.
We assume that a classical particle with a charge $e$ is scattered $n$ times
at $x_i^\mu$ ($i = 1,2,...,n$) and
changes its momentum from $p_{i-1}^\mu$ to $p_i^\mu$ by each scattering.
The current density in that process is calculated from
\beq
j^\mu (x) = e \int \dd t \frac{\dd y^\mu(t)}{\dd t} \delta^4 \lmk x - y(t) \rmk,
\eeq
where $t$ is the time variable.
The trajectory $y(t)$ is written as
\beq
 y^\mu (t) = x_{i}^\mu + \frac{p^\mu_i}{p_i^0} (t-x_i^0) \qquad \text{for } x^0_{i} < t < x^0_{i+1}.
\eeq
The Fourier transform of the current density is thus given as
\beq
j^\mu (k) 
= i e \sum_{i=1}^n e^{i k x_i}
\lmk \frac{p_i^\mu}{k p_i} - \frac{p_{i-1}^\mu}{k p_{i-1}} \rmk.
\label{Fourier trans of current}
\eeq
The spectrum of photons radiated during scatterings is calculated from
\beq
 \frac{\dd^3 n_\gamma}{\dd k^3} = - \frac{\left\vert j(k) \right\vert^2}{2k^0 (2 \pi)^3}.
\eeq
The incoherent limit $k (x_i-x_j) \gg 1$ corresponds to the usual Bethe-Heitler limit,
in which each scattering can be regarded as an independent inelastic scattering process.
On the other hand,
in the limit of $k (x_i-x_j) \ll 1$,
adjacent terms in Eq.~(\ref{Fourier trans of current}) are canceled with each other
and the radiations are suppressed.
This is a physical origin of the LPM effect.
The LPM effect is thus interpreted as an interference effect between a parent particle and a daughter particle
which is emitted collinearly.
Although we consider the case of classical electrodynamics as an illustration,
it has been proven that the same suppression effect is realized in quantum field theories, including QED and QCD~\cite{Landau-Pomeranchuk, Migdal,
Gyulassy:1993hr,
Arnold:2001ba, Arnold:2001ms, Arnold:2002ja, Besak:2010fb}.

When we write the position vector of a parent particle as $x^\mu = \lmk \Delta t, \Delta t \hat{z} \rmk$,
the interference effect
remains until the phase factor varies significantly as%
\footnote{
	In the last equation,
	we assume that the angle $\theta$ varies dominantly by the change of direction of the daughter particle.
	However, since the parent particle also changes its direction due to the elastic scatterings,  
	it  contributes to the angle as $\theta \simeq p_\perp / p^0$.
	In fact, in the case of photon emissions, for example, this effect dominates the time scale of LPM suppression, and 
	it is given by
	\begin{align*}
		\Delta t \sim \frac{1}{\alpha T} \left( \frac{E_i^2}{k^0 T}  \right)^{1/2}.
	\end{align*}
	However,
	the conclusion in this subsection and calculations in the subsequent sections are unchanged even in this case,
	because the parent particle similarly loses its energy
	dominantly through a splitting into daughter particles with $k^0 \sim E_i/2$.
}
\beq
 1 \lesssim
 k x \sim \Delta t k^0 \theta^2 \sim \Delta t k^2_{\perp} / k^0,
\eeq
where
$k_\perp$ is the perpendicular momentum of the daughter particle 
and $\theta$ ($=k_\perp / k^0$) is the emission angle of the daughter particle.
Subsequent inelastic scattering processes are suppressed until this condition is satisfied,
and thus, the inelastic scattering rate per daughter momentum
is suppressed by a factor of
$1/n_{\rm min} \sim 1/ \Delta t \Gamma_{\rm el}$,
where
$n_{\rm min}$ is the lowest number of elastic scatterings to avoid the interference effect.
In summary, the inelastic scattering rate
is determined as
\beq
 \Gamma_{\rm inelastic} \sim \min \lkk \la \sigma_{\rm inelastic} n \ra, \int \frac{\dd k^0}{k^0} \frac{\alpha}{\Delta t (k^0)} \rkk,
 \label{inela rate}
\eeq
where $\Delta t (k^0) \sim k^0 / k_\perp^2$.
The first and second terms in this equation correspond to the limit of $k (x_i-x_j) \gg 1$ and $k (x_i-x_j) \ll 1$, respectively.
This is the correct inelastic scattering rate
with the LPM effect taken into account.

We need to estimate $\Delta t$ ($\sim k^0 / k_\perp^2$) in order to determine the inelastic scattering rate given in Eq.~(\ref{inela rate}).
If we could neglect subsequent scatterings for the daughter particle,
its perpendicular momentum is given as $k_\perp \sim \alpha^{1/2} T$.
In this case, $\Delta t$ is given as
\beq
 \Delta t \sim \frac{k^0}{\alpha T^2}.
\label{del t 1}
\eeq
When we take account of soft elastic scatterings for the daughter particle,
its perpendicular momentum evolves as random walk and is described as
\beq
 \lmk \Delta k_\perp \rmk^2 \sim \qel t,
\eeq
where $\qel$ is a diffusion constant written by the soft elastic scattering rate for the daughter particle
as
\beq
 \qel \sim \int \dd^2 q_\perp \frac{\del \Gamma_{\rm el}}{\del q_\perp^2} q_\perp^2 \sim \alpha^2 T^3.
\eeq
Using these equations, we obtain
\beq
 \Delta t \sim \lmk \frac{k^0}{\qel} \rmk^{1/2} \sim \frac{1}{\alpha T} \lmk \frac{k^0}{T} \rmk^{1/2}.
 \label{del t 2}
\eeq
Since $\Delta t$ in Eq.~(\ref{del t 1}) is larger than the one in Eq.~(\ref{del t 2}),
the latter one determines the time when
the LPM effect becomes irrelevant.
The rate of inelastic scatterings is therefore determined by Eqs.~(\ref{inela rate}) and (\ref{del t 2}).

Taking into account the LPM effect,
we obtain the rate of energy loss through inelastic scattering processes as
\beq
 \left. \frac{\dd E}{\dd t} \right\vert_{\rm inelastic} \sim \int^{E_i/2} \dd k^0 \frac{\alpha}{\Delta t (k^0)} \sim
 \alpha^2 T^2 \sqrt{\frac{E_i}{T}}.
\eeq
Since this rate is larger than
the rate of energy loss through elastic scatterings given by Eq.~(\ref{elastic dEdt}) for high-energy particles with $E_i \gg T$,
they lose their energy
mainly by inelastic scatterings.
Note that
the energy loss rate of inelastic scatterings per daughter momentum
is larger for larger daughter momentum.
Therefore, high-energy particles most efficiently lose their energy by a splitting into two particles with the energy
of order $E_i/2$.
The daughter particles continue to split and their number density grows exponentially.

%%%%%%%%%%%%%%%%%%%%%%%%%%%%%%%%%%%%%%%%%%%%%%%%%%%%%%%%%%%%%%%%
\subsection{\label{history}Thermal history}
%%%%%%%%%%%%%%%%%%%%%%%%%%%%%%%%%%%%%%%%%%%%%%%%%%%%%%%%%%%%%%%%

In this subsection, we briefly explain the evolution of the thermal plasma
during the reheating process
using the scattering rate derived in the previous subsection.
For a more detailed discussion, see Ref.~\cite{Harigaya:2013vwa}
(see also Refs.~\cite{Kurkela:2011ti,Davidson:2000er,Arnold:2002zm,Jaikumar:2002iq}).

After inflation,
the energy density of the Universe is dominated by
an oscillating inflaton
and decreases as $a^{-3}$, where $a$ is the scale factor of the Universe.
The inflaton decays into radiation, which is a starting point of reheating of the Universe.
Let us write the mass and the decay rate of the inflaton as $\mphi$ and $\Gphi$, respectively.
Daughter particles produced from inflaton decay have very high energy of the order of $\mphi$,
and the number density of them, $n_h$, is given as
\beq
 n_h(t) \simeq \int_t \dd t' \ n_\phi(t') \Gamma_\phi \sim n_\phi(t) \Gamma_\phi t \sim \frac{\Gamma_\phi \Mpl^2}{\mphi t},
\eeq
for $\Gamma_\phi \ll H$, where $H$ is the Hubble parameter.
At the early stage of reheating,
inelastic scatterings between high-energy particles generate many low-energy particles 
almost without losing their energy.
Soon after that, low-energy particles thermalize by their own interaction, and
the number density of particles in the thermal plasma is larger than that of high-energy particles at the same time.
However, the energy density of radiation is still dominantly stored by high-energy particles with momentum $m_\phi$,
which is not thermalized yet.
Eventually, high-energy particles lose their energy via inelastic scattering processes 
with the thermal plasma, which is the bottleneck process of thermalization in this case.
We define a momentum $\ksplit$ such that
particles with the momentum $\ksplit$
lose their energy completely and are thermalized by the time of $H^{-1}$.
Once a splitting of a momentum $k$ becomes efficient,
a particle with a momentum smaller than $k$ loses its energy rapidly by splitting processes.
Therefore, $\ksplit$ is given by
\beq
 \frac{\dd \Gamma_{\text{inelastic}}}{\dd \log k^0} (\ksplit ) \sim H.
\eeq
High-energy particles efficiently supply their energy into the thermal plasma
by emitting particles with the momentum $\ksplit$.
Since the energy conservation implies $T^4 \sim \ksplit n_h$, we obtain
\beq
 T &\sim& \alpha^4 \lmk \frac{\Gamma_\phi \Mpl^2}{\mphi^3} \rmk \mphi \lmk \mphi t \rmk, \\
 \ksplit &\sim& \alpha^{16} \lmk \frac{\Gamma_\phi \Mpl^2}{\mphi^3} \rmk^3 \mphi \lmk \mphi t \rmk^5.
\eeq
Each high-energy particle completely loses its energy when the splitting momentum
becomes comparable to the maximum momentum: $\ksplit \sim \mphi$.
Thermalisation of high-energy particles is thus completed at the time given as
\beq
 \lmk \mphi t_\text{th} \rmk \sim \alpha^{-16/5} \lmk \frac{\Gamma_\phi \Mpl^2}{\mphi^3} \rmk^{-3/5}.
 \label{t_th}
\eeq
This is the time when the temperature of the Universe is maximum:
\beq
 T_{\text{max}} \sim \alpha^{4/5} \lmk \frac{\Gamma_\phi \Mpl^2}{\mphi^3} \rmk^{2/5} \mphi.
 \label{T_max}
\eeq
Note that the energy density of the Universe is still dominated by that of the inflaton.

Until the Hubble parameter becomes comparable to the decay rate of the inflaton (i.e. $H > \Gphi$),
the energy density of the Universe is still dominated by that of the inflaton.
Thus, we obtain the following approximation during $t \lesssim \Gphi^{-1}$:
\beq
 \rho_\phi &\simeq& \frac{4 \Mpl^2}{3 t^2}, \label{rho_phi evol}\\
 H &\simeq& \frac{2}{3 t}, \label{H evol}\\
 \rho_{\rm r} 
 &\simeq& 
%  \Red{ 
\frac{3}{5} 
 \rho_\phi \Gamma_\phi t \simeq \frac{4 \Gamma_\phi \Mpl^2}{5 t}.
% &\sim& \rho_\phi \Gamma_\phi t \simeq \frac{4 \Gamma_\phi \Mpl^2}{3 t}.
% }
  \label{rho_r evol}
\eeq
After the time of $t_{\rm th}$,
high-energy particles from inflaton decay are thermalized soon, 
and thus, the energy density of
radiation $\rho_{\rm r}$ is simply characterized by the temperature $T$.
From the last relation, we obtain the temperature of radiation as
\beq
 T \simeq \lmk \frac{
% \Red{
 36
% } 
 H \Gphi \Mpl^2}{g_* (T) \pi^2} \rmk^{1/4} \propto a^{-3/8},
 \label{T during reheating}
\eeq
where $g_*$ is the effective number of relativistic degrees of freedom.

We define
reheating temperature $\TR$ as the temperature at which the energy density of the inflaton and radiation
are equal to each other.
The reheating temperature is thus obtained from the equation $H \simeq \Gphi$ and is given as
\beq
 \TR \simeq \lmk \frac{90}{g_* (\TR) \pi^2} \rmk^{1/4}
 \sqrt{\Gamma_\phi \Mpl}.
 \label{T_RH}
\eeq
After the era of reheating,
the energy density of the Universe is dominated by
that of radiation and decreases as $\propto a^{-4}$.

%%%%%%%%%%%%%%%%%%%%%%%%%%%%%%%%%%%%%%%%%%%%%%%%%%%%%%%%%%%%%%%%
\section{\label{production}DM production mechanisms}
%%%%%%%%%%%%%%%%%%%%%%%%%%%%%%%%%%%%%%%%%%%%%%%%%%%%%%%%%%%%%%%%

In this section, we discuss DM production for 
a theory with
the inflaton mass $\mphi$, the  WIMP (DM) mass $\mDM$ ($\ll \mphi$), and a low reheating temperature $\TR$ ($\ll \mDM$).
We explain three mechanisms to produce DM:
thermal production (Sec.~\ref{thermal}),
production through a cascade shower from inflaton decay (Sec.~\ref{shower}),
and production through inelastic scatterings between high-energy particles and the thermal plasma (Sec.~\ref{inela}).
These mechanisms are additional contributions with each other, and thus, 
the predicted DM density is the sum of these contributions in a scenario with low reheating temperature.

%%%%%%%%%%%%%%%%%%%%%%%%%%%%%%%%%%%%%%%%%%%%%%%%%%%%%%%%%%%%%%%%
\subsection{\label{thermal}Thermal production}
%%%%%%%%%%%%%%%%%%%%%%%%%%%%%%%%%%%%%%%%%%%%%%%%%%%%%%%%%%%%%%%%

In this subsection, we explain thermal production of DM in the Universe with low reheating temperature.
Even if $\TR \ll \mDM$,
DM is generated thermally during the inflaton-dominated era~\cite{Chung:1998rq, Giudice:2000ex}.
The condition to generate DM thermally is given as $T_{\text{max}} \gtrsim \mDM$, where $T_{\text{max}}$ is
the maximum temperature of the Universe
derived as Eq.~(\ref{T_max}) and is rewritten in terms of $\TR$ and $\mphi$ as
\beq
 T_{\text{max}} \sim \lmk \alpha^2 \frac{\TR^2 \Mpl}{\mphi^3} \rmk^{2/5} \mphi,
\eeq
where we use Eq.~(\ref{T_RH}) and omit $O(1)$ factors.
If $T_{\text{max}} \ll \mDM$,
the DM density is exponentially suppressed.
In the following, we calculate the DM density for the case of $T_{\rm max} \gtrsim \mDM$ ($\gg \TR$).

As is the case with typical WIMP scenarios, we assume that DM has an odd $Z_2$ parity and thus is stable
and has the weak interaction.%
\footnote{
If DM interacts with Standard Model particles 
only through a heavy mediator or a higher dimensional interaction
so that DM has never been in thermal equilibrium,
DM is non-thermally produced mainly around the end of the reheating~\cite{Moroi:1993mb, Bolz:2000fu, 
Pradler:2006qh, Kusenko:2010ik, Mambrini:2013iaa}.
}
We express its thermal-averaged annihilation cross section as
\beq
 \la \sigma_{\rm ann} v \ra \equiv \frac{\alpha_{\rm w}}{\mDM^2} \lmk c_s + \frac{T}{\mDM} c_p \rmk,
\eeq
where $\alpha_{\rm w}$ is the fine-structure constant of the weak interaction.
The terms with the coefficients $c_s$ and $c_p$ describe the $s$-wave and $p$-wave contributions
in a non-relativistic expansion of the cross section.

The number density of DM decreases through the annihilation and the Hubble expansion.
Since the rate of the annihilation is proportional to the number density of DM,
the annihilation process becomes irrelevant and the number density freezes out
when the following condition is satisfied:
\beq
 n_{\rm DM}^{eq} \lmk T_{\rm F} \rmk \la \sigma_{\rm ann} v \ra \simeq H \lmk T_{\rm F} \rmk,
 \label{freeze-out condition}
\eeq
where $n_{\rm DM}^{eq}$ is the number density of DM with the assumption of thermal equilibrium. 
Since we consider the case of $\TR \ll \mDM (\sim T_{\rm F})$,
DM decouples from the thermal plasma during the inflaton-dominated era,
in which the temperature of the thermal plasma obeys Eq.~(\ref{T during reheating}).
Defining $x_{\rm F} \equiv \mDM / T_{\rm F}$, we rewrite the condition (\ref{freeze-out condition}) as
\beq
 x_{\rm F} \simeq \log \lkk \frac{6}{\sqrt{5} \pi^{5/2}} \frac{g_*^{1/2} (\TR)}{g_* (T_{\rm F})} \frac{\Mpl}{\mDM}
 \alpha_{\rm w} \lmk c_s + \frac{5}{4} c_p x_{\rm F}^{-1} \rmk x_{\rm F}^{1/2} \frac{\TR^2}{T_{\rm F}^2} \rkk. 
 \label{x_F}
\eeq
The DM freeze-out occurs earlier compared with the ordinary case of thermal production of DM
roughly by a factor of $\log [ \TR^2 / T_{\rm F}^2 ]$.
This is because
the energy density of radiation is less than that of inflaton during the inflaton-dominated era
and the expansion rate of the Universe evolves faster than in the ordinary case.
%\Red{
We obtain the abundance of DM as
\beq
  \left. n_{\rm DM} \right\vert_{T = T_{\rm F}}
  \simeq  
 \frac{2 \mDM^3}{  \lmk 2\pi x_{\rm F} \rmk^{3/2}} e^{- x_{\rm F} }. 
\eeq
%}

%\Red{
Here we comment on the case of $\mDM \gtrsim ( \TR^2 \Mpl )^{1/3}$. 
In this case, 
DM never reaches the chemical equiliblium 
because 
the combination of $n_{\rm DM}^{eq} \lmk T \rmk \la \sigma_{\rm ann} v \ra$ 
is always less than the Hubble parameter, $H(T)$. 
Since we assume $T_\text{max} \gtrsim \mDM$, 
DM is produced through a pair creation ($=$ inverse annihilation) process. 
Assuming its cross section to be $\alpha_{\rm w} T^{-2}$ for $T \gtrsim \mDM$, 
we find that DM is dominantly produced at $T = \mDM$ and 
obtain its abundance as 
\beq
 \left. n_{\rm DM} \right\vert_{T = \mDM}
  &\simeq& 
  \frac{ \lmk n_{\rm DM}^{\rm eq} \rmk^2_{ T=\mDM} \la \sigma_{\rm ann} v \ra}
  {H(T= \mDM)}, \\
  &\simeq&  
  \frac{9 \zeta^2 (3) }{\sqrt{10} \pi^5} \frac{\alpha_{\rm w} g_*^{1/2} \lmk \TR \rmk}{g_* \lmk \mDM \rmk } 
  \Mpl \TR^2 
  \quad \text{ for } \  \mDM \gtrsim \lmk \TR^2 \Mpl \rmk^{1/3}, 
\eeq
where $\zeta(3) \simeq 1.20205\dots$ is the Riemann zeta function. 
%}

The present energy density of DM from the thermal production
divided by the entropy density of the Universe is thus given as
\beq
 \left. \frac{\rho_{\rm DM}^{\rm th}}{s} \right\vert_{\text{now}}
 &\simeq& \left. \frac{3\TR \rho_{\rm DM}^{\rm th}}{4\rho_\phi} \right\vert_{\text{RH}}, \\
 &\simeq& \left. \frac{3\TR \rho_{\rm DM}^{\rm th}}{4\rho_\phi} \right\vert_{\text{F}}, \\
 &\simeq& \left. \frac{\rho_{\rm DM}^{\rm th}}{s} \right\vert_{\text{F}} \lmk \frac{\TR \rho_{\rm r}}{T_{\rm F} \rho_\phi} \rmk_{\rm F}, \\
 &\simeq& \left. \frac{\rho_{\rm DM}^{\rm th}}{s} \right\vert_{\text{F}} \lmk \frac{\TR}{T_{\text{F}}} \rmk^{5},
 \label{DM density1}
\eeq
where the subscripts ``RH'' and ``F'' represent the corresponding value at the time of reheating and DM freeze-out, respectively.
We use $s = 4 \rho_{\rm r} / 3 T$ in the first and third lines, $\rho_{\rm DM}^{\rm th} \propto \rho_\phi \propto a^{-3}$ in the second line,
and $T \propto a^{-3/8}$ in the last line.
The DM abundance is suppressed compared with the ordinary case of thermal production of DM
due to the entropy production from the inflaton decay after the time of DM freeze-out.
This scenario has been considered in the literature
in order to suppress the abundance of WIMPs with relatively large mass~\cite{Chung:1998rq, Giudice:2000ex}.

%%%%%%%%%%%%%%%%%%%%%%%%%%%%%%%%%%%%%%%%%%%%%%%%%%%%%%%%%%%%%%%%
\subsection{\label{shower}DM production through cascade shower from inflaton decay}
%%%%%%%%%%%%%%%%%%%%%%%%%%%%%%%%%%%%%%%%%%%%%%%%%%%%%%%%%%%%%%%%

DM may be directly produced by the decay of the inflaton~\cite{Moroi:1994rs, Kawasaki:1995cy, Moroi:1999zb, Gelmini:2006pw}.
The number density of DM from this contribution at the temperature $T=\TR$ is given as
\beq
 \left. n_\DM^{\text{dir}} \right\vert_{T=\TR} = \Br\lmk \phi \to \DM \rmk \left. n_{\phi} \right\vert_{T=\TR},
 \label{n_dir}
\eeq
where $n_{\phi}$ is the number density of the inflaton.
We denote the branching ratio of inflaton decay into DM as $\Br ( \phi \to \text{DM})$,
which depends on the model one considers.
For example, $\Br ( \phi \to \text{DM}) = O(1)$ in SUSY theories due to SUSY and
the R-parity conservation.
From Eq.~(\ref{n_dir}), one may estimate the DM abundance from the decay of the inflaton at the present time as
\beq
 \left. \frac{n_\DM^{\text{dir}}}{s} \right\vert_{\text{now}} &\simeq& \left. \TR \frac{3 n_\DM^{\text{dir}}}{4 \rho_\phi} \right\vert_{T=\TR}, \\
 &\simeq& \frac{3 \TR}{4 \mphi} \Br\lmk \phi \to \DM \rmk.
\eeq
However,
we have to take account of the contribution from the cascade decay of the inflaton.
This has been investigated in Ref.~\cite{Kurata:2012nf}, where they assume the minimal SUSY standard model.
Using generalized DGLAP equations~\cite{Sarkar:2001se, Barbot:2002ep},
they have found that more than one DM (LSP, in that paper) is produced
through each cascade decay of the inflaton.
Their results are written as
\beq
 \left. \frac{n_\DM^{\text{shower}}}{s} \right\vert_{\text{now}} &\simeq& \frac{3 \TR}{4 \mphi} \sum_i \Br\lmk \phi \to i \rmk \nu_i,
\eeq
where $\nu_i$ is the averaged number of DM in a shower produced by a primary particle $i$.
The factor $\nu_i$ increases with increasing the mass of the inflaton
and strongly depends on particle species $i$.
For example, if $\mphi = 10^{13}$ GeV and $\mDM = 1$ TeV,
$\nu_i$ is calculated as $\mathcal{O}(1), \mathcal{O}(10^{2})$, and $\mathcal{O}(10)$
for (s)neutrinos, ${\rm SU}(3)_{\rm c}$-charged particles, and the other particles, respectively.%
\footnote{
These results have been obtained by extrapolating data points
of the inflaton mass $\mphi \le 10^{10}$ GeV, and thus, there are some uncertainties.
}

%%%%%%%%%%%%%%%%%%%%%%%%%%%%%%%%%%%%%%%%%%%%%%%%%%%%%%%%%%%%%%%%
\subsection{\label{inela}DM production through inelastic scatterings}
%%%%%%%%%%%%%%%%%%%%%%%%%%%%%%%%%%%%%%%%%%%%%%%%%%%%%%%%%%%%%%%%

In this section, we consider inelastic scattering processes between high-energy particles and the thermal plasma.
Since relevant processes are inelastic scatterings into two high-energy particles,
as explained in Sec.~\ref{thermalization},
we refer to those processes as splittings.
We concentrate on the time when the temperature is in the interval $\TR < T \ll T_{\rm max}$, i.e., $\Gphi^{-1} \gtrsim t \gg t_{\rm th}$.

The inflaton decays into particles with the energy of the order of its mass $\mphi$,
and the daughter particles lose their energy by splitting continuously.
DM is produced with a certain rate
throughout these splitting processes
when the energy of the splitted particles is sufficiently large.
We define a threshold energy as
\beq
 \Eth \equiv \frac{\mDM^2}{4 T}.
 \label{Eth}
\eeq
When a high-energy particle has energy larger than this threshold energy,
inelastic scatterings between the high-energy particle and the thermal plasma can produce DM.
The cross section of a DM production process is suppressed by the mass of DM as~\cite{Allahverdi:2002nb}
%\Red{
%%
\beq
 \sigma_{\rm DM} \sim \max \lkk \frac{\alpha_{\text{DM}}^2}{s}, \frac{\alpha_{\text{DM}}^3}{\mDM^2} \rkk,
\eeq
%%
%}
where $s$ is one of the Mandelstam variables and is given by $4 E T$.
The former cross section is nothing but the ordinary pair creation 
from an
annihilation of the high-energy particle and a particle in the thermal plasma.
The physics behind the latter cross section is equivalent to the $e^+ e^-$ pair production from a high-energy photon
interacting with a nuclei,
where the cross section
is proportional to
the inverse of the squared electron mass.
%\Red{
We write the fine-structure constant of DM production processes as $\alpha_\text{DM}$, 
which is generally different from the one appearing in the inelastic scattering rate 
(see Eqs.~(\ref{inela rate}) and (\ref{del t 2})). 
%}
For a reaction with energy just above the threshold (i.e. $E \gtrsim \Eth$), which we are most interested in as explained below,
the rate of the DM production process is given as
%%
%\Red{
\beq
 \left. \Gamma_{\rm DM} \right\vert_{E \sim \Eth} \sim \la \sigma_{\rm DM} n \ra \sim \frac{\alpha_{\text{DM}}^2 T^3}{\mDM^2}.
\label{Gamma_DM}
\eeq
%}
%%
Note that this is so small that the LPM effect is irrelevant for this process
(see Eqs.~(\ref{inela rate}) and (\ref{del t 2})),
and thus, the rate of the DM production process is indeed given by this formula.

Here we estimate the number density of DM produced through inelastic scattering processes.
A more detailed discussion is done in the Appendix, where we solve the Boltzmann equation
describing inelastic scattering processes.
Let us consider the evolution of high-energy particles.
First, they are produced by the decay of the inflaton and have the energy of the order of its mass $\mphi$.
Soon after that, the daughter particles split into many high-energy particles.
The high-energy particles continue to split, and their number density grows exponentially.
Given a certain time when their energy is of the order of $E$,
we can estimate
their number density $n_h$ as
\beq
\label{number high}
 n_h \sim \frac{\mphi}{E} n_\phi 
 \Gphi t \qquad \text{for} \quad  t_{\rm th} \ll t \lesssim \Gphi^{-1},
\eeq
from the conservation of energy.
Here we use the fact that the splitting process is much faster than the decay of the inflaton
since they satisfy the inequality $\Gsplit \gg \Gphi$ for $t_{\rm th} \ll t \lesssim \Gphi^{-1}$.
Throughout these processes,
DM is also produced
by scatterings of high-energy particles
with the rate given in Eq.~(\ref{Gamma_DM}),
until they lose their energy down to $\Eth$.
At a certain time when their energy is of the order of $E$,
the number density of DM 
that is produced during the splitting of the high-energy particles (i.e.~$\Gamma_\text{inelastic}(E)^{-1}$)
is therefore given by
\beq
 n_{\rm DM}^{\rm sca} \sim \frac{\Gamma_{\rm DM}}{\Gamma_{\rm inelastic}} n_h 
 \sim 
%\Red{
 \frac{\alpha_{\text{DM}}^2 T^3}{\mDM^2} \frac{\sqrt{E}}{\alpha^2 T \sqrt{T}} \frac{\mphi}{E} n_\phi
 \Gphi t \qquad \text{for} \quad  t_{\rm th} \ll t \lesssim \Gphi^{-1}.
%}
 \label{nDM}
\eeq
Equations~(\ref{T during reheating}) and (\ref{nDM}) imply that the abundance of DM
increases with decreasing $E$ and $T$.
Taking into account the inequality $\mphi/2 \ge \Eth$, which constrains
the temperature as $T \ge \mDM^2 / 2\mphi$ to produce the DM,
we conclude that the energy density of the DM is given by
%%
%\Red{
\beq
\label{entropy ratio}
 \frac{\rho_{\rm DM}^{\rm sca}}{s}
 &\sim& 
 \left.
 \mDM 
 \frac{\alpha_{\text{DM}}^2 T^3}{\mDM^2} \frac{\sqrt{\Eth}}{\alpha^2 T \sqrt{T}} 
 \frac{\mphi}{\Eth}
 \Gphi t \right\vert_{T = {\rm max}(\TR, \mDM^2/2\mphi)}\times
 \left. \frac{n_\phi}{s} \right\vert_{T=\TR},
 \\
 &\sim& 
 \left\{
 \begin{array}{ll}
 \displaystyle{\frac{\alpha_{\text{DM}}^2}{\alpha^2} \frac{\TR^3}{\mDM^2}} &\qquad \text{for} \quad \displaystyle{\mphi \ge \frac{\mDM^2}{2 \TR}}, \vspace{3mm}\\
 \displaystyle{\frac{\alpha_{\text{DM}}^2}{\alpha^2} 
 \frac{4\TR^5 \mphi^2}{\mDM^6}} &\qquad \text{for} \quad 
 \displaystyle{\frac{\mDM^2}{2 T_{\rm max}} \ll \mphi < \frac{\mDM^2}{2\TR}},
  \end{array}
  \right.
 \label{main result}
\eeq
%}
%%
where we use Eq.~(\ref{T during reheating}) to express $t$ in terms of $T$ as $\Gphi t \sim \TR^4 / T^4$.
The first equality is justified by solving the Boltzmann equation in the Appendix.
We should emphasize that
this result is independent of the mass of the inflaton
once the condition to produce the DM is satisfied
at $T=\TR$ (i.e. $\mphi \ge \mDM^2 / 2\TR$).

In the above analysis, we assume that
the thermalization of decay products proceeds through inelastic scatterings and elastic scatterings are negligible,
which is the case when decay products have gauge interactions as we  have shown in Sec.~\ref{energy loss}.
However, when the temperature of the thermal plasma is smaller than the QCD scale, $\Lambda_{\rm QCD}=O(100)$ MeV,
${\rm SU}(3)_{\rm c}$-charged particles hadronize.
Some hadrons, such as the neutral pion, have no gauge interactions, 
and the energy loss by splitting processes is suppressed.
Therefore, the energy loss by elastic processes is important, and
the number density of high-energy particles given in Eq.~(\ref{number high})
is over-estimated for $T\lsim\Lambda_{\rm QCD}$.
Since the estimation of the number density of high-energy particles for $T\lsim\Lambda_{\rm QCD}$
suffers from large uncertainties due to the non-perturbative feature of hadronization,
we leave it for a future work.

%%%%%%%%%%%%%%%%%%%%%%%%%%%%%%%%%%%%%%%%%%%%%%%%%%%%%%%%%%%%%%%%
\subsection{\label{subsec3-4}Summary}
%%%%%%%%%%%%%%%%%%%%%%%%%%%%%%%%%%%%%%%%%%%%%%%%%%%%%%%%%%%%%%%%

The amount of DM at the present time has been observed by
the $Planck$ collaboration~\cite{Ade:2013zuv} as
\beq
 \Omega_{\rm DM} h^2 = 0.1196 \pm 0.0031.
\eeq
Using $\Omega_{\rm DM} h^2 \simeq (\rho_{\rm DM} /s)/3.5$ eV and Eq.~(\ref{main result}),
we obtain the relation between the mass of DM and the reheating temperature as
\beq
 \mDM \sim 1.5 \text{ TeV} 
%\Red{
 \lmk \frac{\alpha_{\text{DM}}}{\alpha} \rmk
%}
 \lmk \frac{\TR}{100\text{ MeV}} \rmk^{3/2},
 \label{result mDM}
\eeq
once the condition of $\mphi \ge \mDM^2 / 2 \TR$ is satisfied and the contribution from the decay of the inflaton is neglected.

Let us discuss whether the annihilation of DM is negligible or not~\cite{FH}
for the parameter of the interest given in Eq.~(\ref{result mDM}).
The annihilation of DM is irrelevant when the following condition is satisfied:
\beq
 \frac{n_{\rm DM} \la \sigma_{\rm ann} v \ra}{H} \ll 1.
\eeq
In the case of $\mphi \ge \mDM^2 / 2 \TR$,
the DM abundance is determined at $T= \TR$, and hence, 
the left-hand side of this inequality should be calculated at $T=\TR$,
and we obtain an upper bound on the abundance of DM as
\beq
 \Omega_{\rm DM} h^2
 &\ll& \frac{1}{3.5 \text{ eV}}
 \lmk \frac{45}{8 \pi^2 g_* (\TR)} \rmk^{1/2}
 \frac{\mDM}{\la \sigma_{\rm ann} v \ra \Mpl \TR}, \\
 &\simeq&
 10^2
 \lmk \frac{10}{g_* (\TR)} \rmk^{1/2} \lmk \frac{10^{-2} \mDM^{-2}}{\la \sigma_{\rm ann} v \ra} \rmk
 \lmk \frac{\mDM}{1.5 \ \TeV} \rmk^3 \lmk \frac{100 \ \text{MeV}}{\TR} \rmk.
 \label{annihilation cond.}
\eeq
For $\mDM$ and $\TR$ given in Eq.~(\ref{result mDM}),
the upper bound on the DM abundance is larger than the observed DM abundance
as long as $\mDM > O(100)$ GeV.%
\footnote{
Here, we implicitly assume that DM loses its momentum just after they are produced.
If that is not the case, the annihilation cross section of DM is as small as $\alpha^2 \Eth^{-2} \ll 10^{-2} m_{\rm DM}^{-2}$,
and the upper bound on DM abundance can be much larger than the reference value given in Eq.~(\ref{annihilation cond.}).
}
Therefore, the prediction in Eq.~(\ref{result mDM}) is valid for $\mDM > O(100)$ GeV.

\begin{figure}[t]
\centering % \begin{center}/\end{center} takes some additional vertical space
\includegraphics[width=.45\textwidth, bb=0 0 360 357]{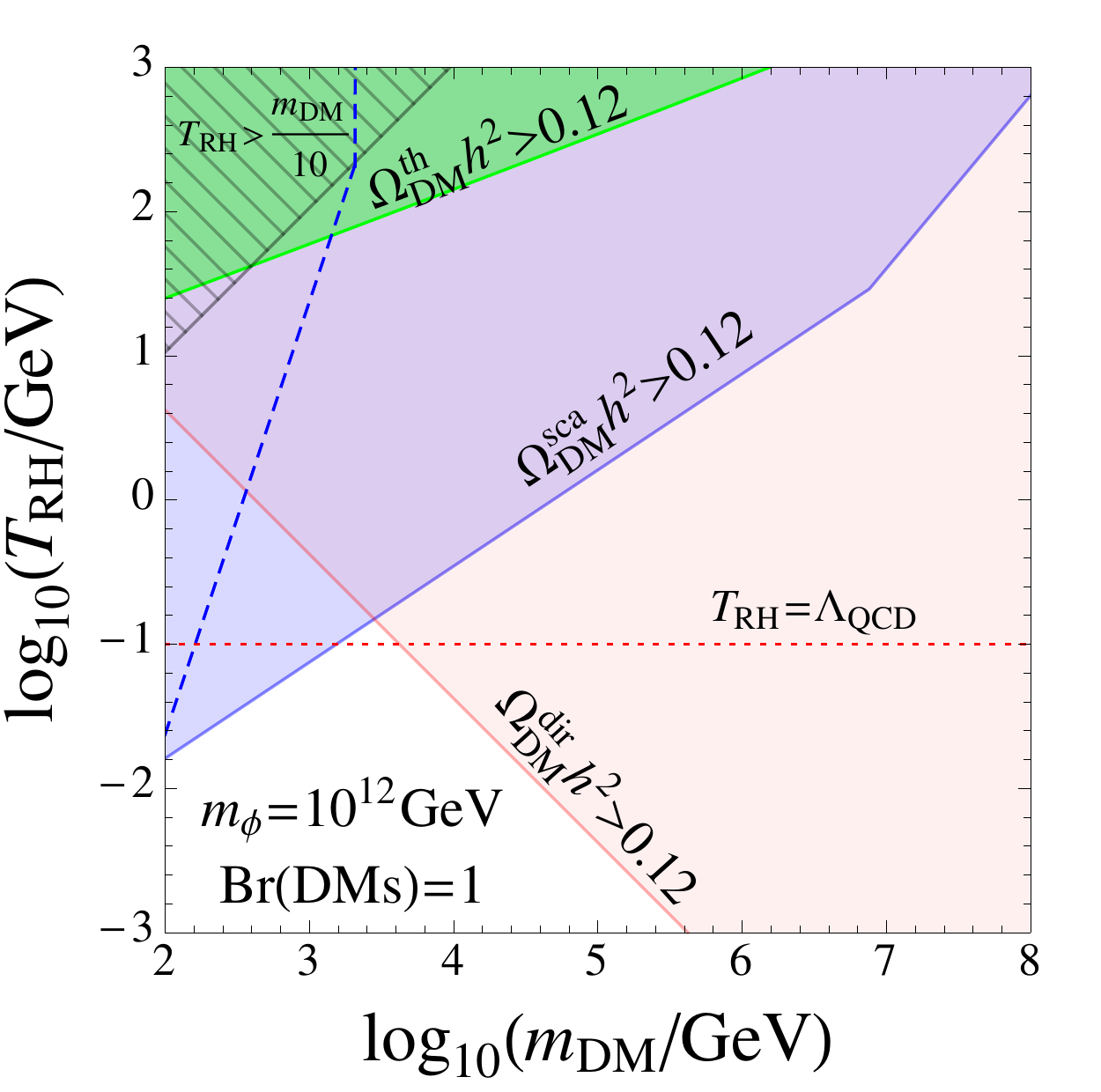} 
\hfill
\includegraphics[width=.45\textwidth, bb=0 0 360 357]{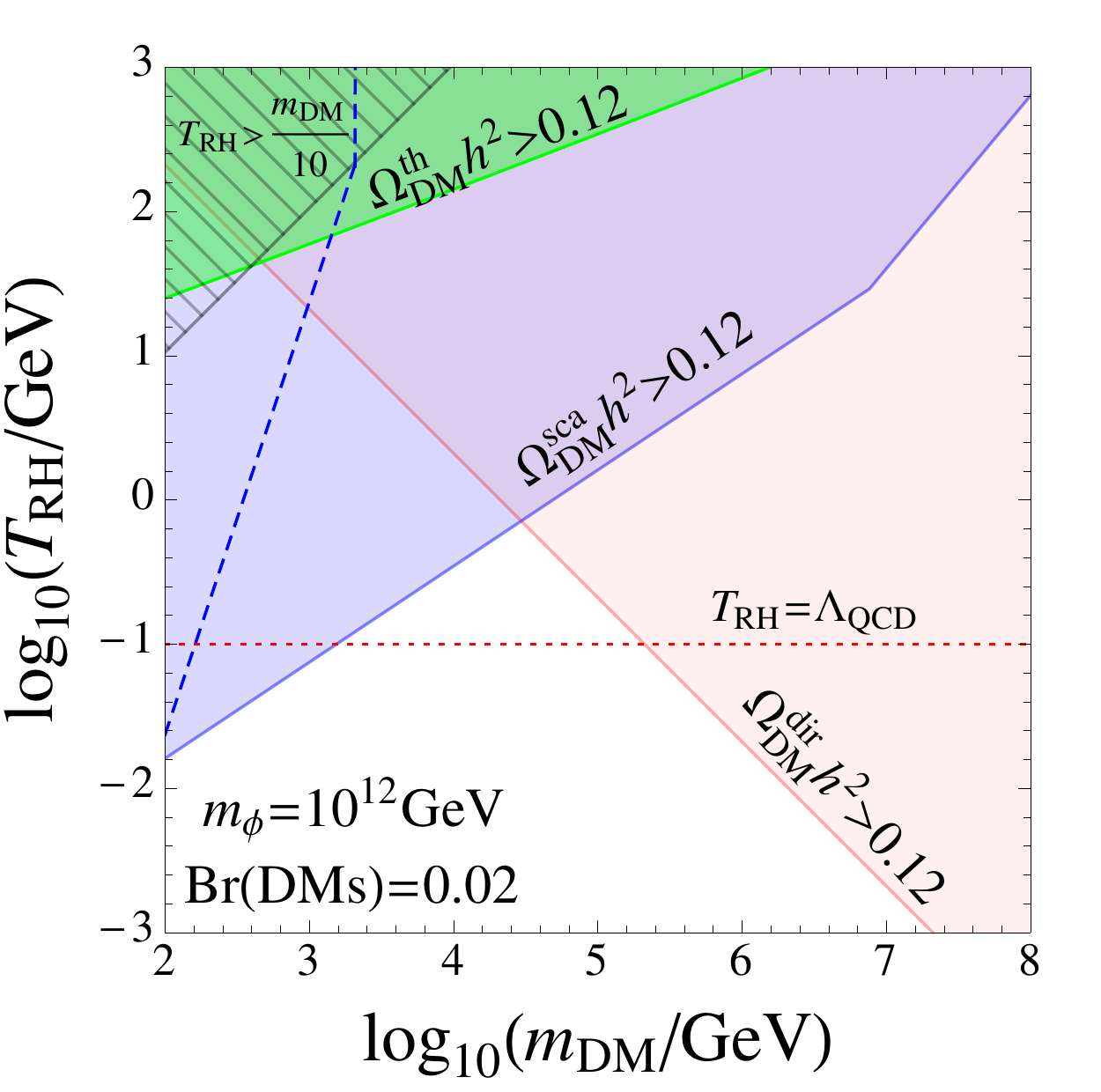}
\caption{
Exclusion plot in a scenario with low reheating temperature.
We assume that the mass of the inflaton $\mphi$ is $10^{12}$ GeV 
and that the branching of inflaton decay into DM is $1$ (left panel) and $0.02$ (right panel).
%\Red{
We also assume
$\alpha_\text{DM} = \alpha$.
%}
The abundance of DM produced from
thermal production ($\Omega_{\rm DM}^{\rm th} h^2$),
direct decay of inflaton ($\Omega_{\rm DM}^{\rm dir} h^2$),
and inelastic scatterings ($\Omega_{\rm DM}^{\rm sca} h^2$)
is larger than that observed in the green, red, and blue shaded regions, respectively.
The striped regions are $\TR > \mDM /10$, in which DM is produced only thermally.
%\Red{
The abundance of DM is less than that observed 
above the blue dashed line due to its annihilation. 
Here, we have assumed that the annihilation of DM is efficient and 
its cross section is $10^{-2} \mDM^{-2}$. 
%}
The red dotted lines represent the reheating temperature
below which $\Omega_{\rm DM}^{\rm sca}h^2$ is over-estimated. 
}
  \label{fig3}
\end{figure}

\begin{figure}[t]
\centering % \begin{center}/\end{center} takes some additional vertical space
\includegraphics[width=.45\textwidth, bb=0 0 360 357]{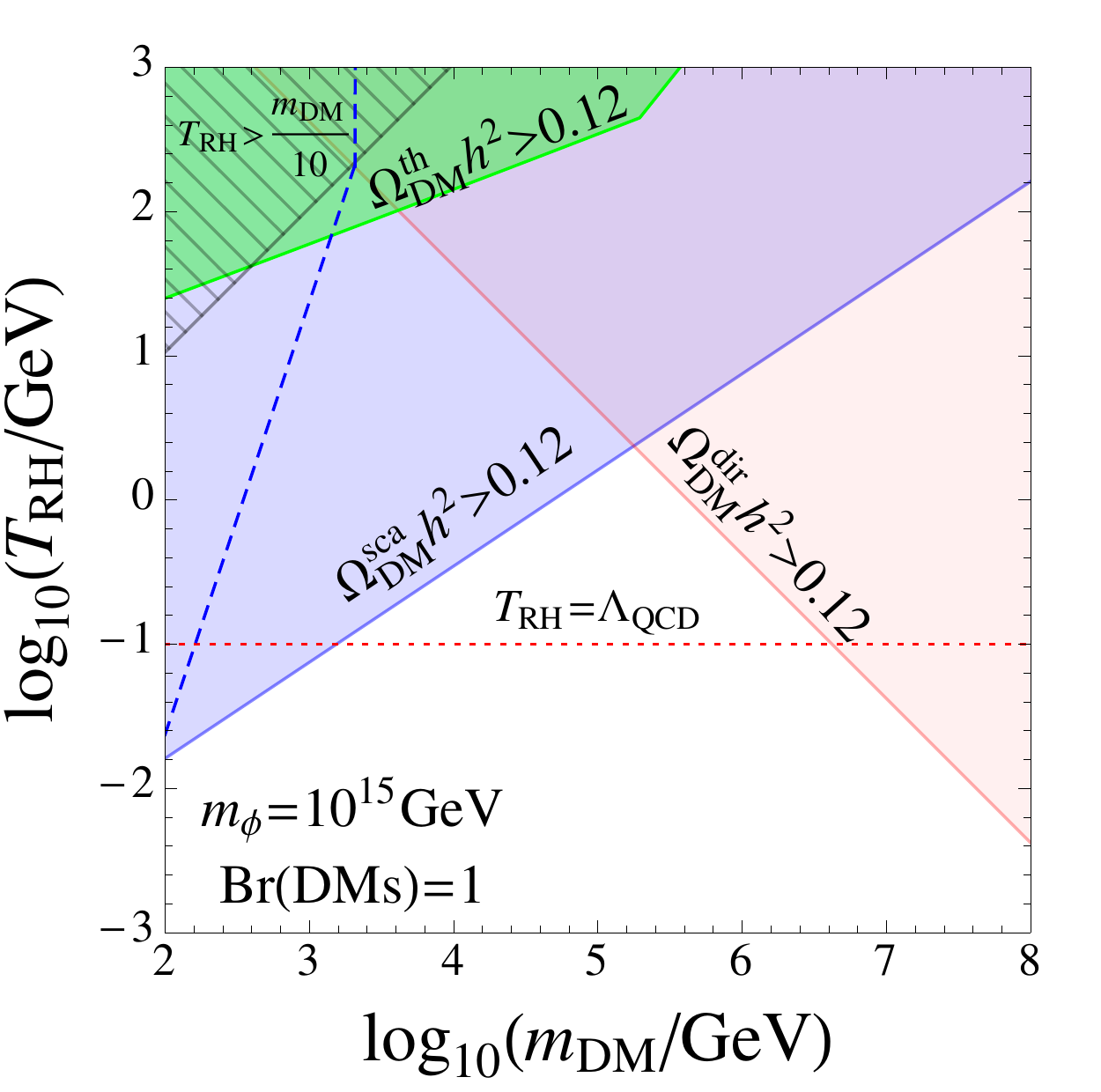} 
\hfill
\includegraphics[width=.45\textwidth, bb=0 0 360 357]{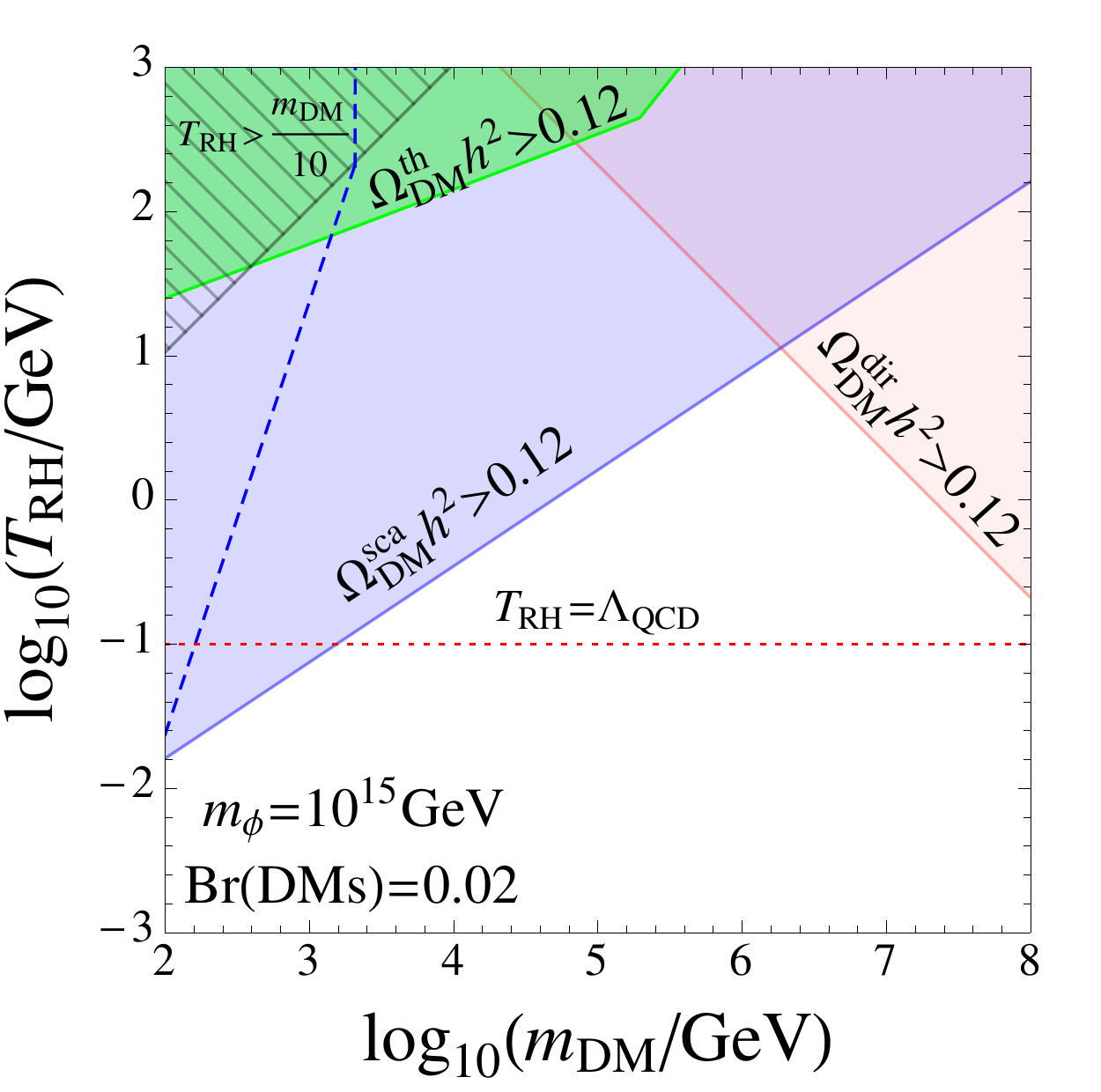}   
\caption{
Same as Fig.~\ref{fig3}, but assuming the mass of the inflaton to be $10^{15}$ GeV.
}
  \label{fig4}
\end{figure}

Figures~\ref{fig3} and \ref{fig4} summarize the results obtained in this section.
Although we take account of the production of DM from direct decay of the inflaton,
we omit the contribution of the DM production from a shower of inflaton decay
since it depends models and has uncertainties, as we have mentioned.
We assume that 
%\Red{
$\alpha_\text{DM} = \alpha$ and 
%}
the mass of the inflaton is $10^{12}$ GeV and $10^{15}$ GeV in Figs.~\ref{fig3} and \ref{fig4},
respectively.%
\footnote{
The inflaton mass of $10^{15}$ GeV is possible in models proposed
in Refs.~\cite{Dimopoulos:1997fv,Takahashi:2010ky,Harigaya:2012pg}.
}
The blue shaded areas are regions in which the energy density of DM
produced by inelastic scatterings exceeds the observed one.
The boundaries of the blue shaded regions are thus given by Eq.~(\ref{result mDM}),
once the condition of $\mphi \ge \mDM^2 / 2 \TR$ is satisfied.
In the case of $\mphi \le \mDM^2 / 2 \TR$, which appears in the upper-right regions of Fig.~\ref{fig3},
the abundance of DM is calculated from the second line of Eq.~(\ref{main result}).
Below the red dotted lines, the reheating temperature is smaller than the QCD scale and the DM density is over-estimated (see the comment in the last paragraph of Sec.~\ref{inela}).
Therefore, for $\mDM\lesssim 10^3$~GeV, the correct DM abundance is obtained 
at a certain reheating temperature 
between the red dotted lines and the lower edges of the blue shaded regions.
Given the mass of the inflaton $\mphi$ and the branching ratio of inflaton decay into DM sector
$\Br \lmk \phi \to \text{DM} \rmk$,
we have an upper bound on the mass of DM
above which the amount of DM from direct decay of inflaton is larger than that observed (red shaded regions). 
%\Red{
If the annihilation of DM is efficient and its
cross section is as large as $10^{-2} \mDM^{-2}$, 
the abundance of DM is 
equal to and less than that observed 
on and above 
the blue dashed lines, respectively. 
The blue dashed line in the striped regions
corresponds to the conventional thermal WIMP scenario. 
%}
The DM production from thermal process calculated in Sec.~\ref{thermal}
is always subdominant in these parameter regions.
Note that DM with a mass of $O(1)$ PeV can account for the abundance of DM
if the branching ratio of the inflaton into the DM sector is suppressed
and the reheating temperature is as large as $10$ GeV.

%%%%%%%%%%%%%%%%%%%%%%%%%%%%%%%%%%%%%%%%%%%%%
%%%%%%%%%%%%%%%%%%%%%%%%%%%%%%%%%%%%%%%%%%%%%

%%%%%%%%%%%%%%%%%%%%%%%%%%%%%%%%%%%%%%%%%%%%%
%%%%%%%%%%%%%%%%%%%%%%%%%%%%%%%%%%%%%%%%%%%%%

Finally, we comment on the case in which
the mass of DM and the reheating temperature are 
within the non-shaded regions 
%\Red{
or above the blue dashed lines 
%}
in Figs.~\ref{fig3} and \ref{fig4}.
In this case,
we need other sources of DM
or other DM candidates to account for the observed DM abundance.
The former solution is easily realized by the decay of long-lived matter:
moduli~\cite{Moroi:1994rs, Kawasaki:1995cy, Moroi:1999zb} or Q-ball~\cite{EnMc, FH, FY, Kamada:2012bk},
for example.
Axion, which is introduced by the Peccei-Quinn mechanism~\cite{PQ}, is one of the well-motivated candidates
for the latter solution.

%%%%%%%%%%%%%%%%%%%%%%%%%%%%%%%%%%%%%%%%%%%%%%%%%%%%%%%%%%
\section{
\label{discussions}discussions}
%%%%%%%%%%%%%%%%%%%%%%%%%%%%%%%%%%%%%%%%%%%%%%%%%%%%%%%%%%

In this section, we discuss the relation between our result and some related topics:
the free-streaming velocity of DM,
Affleck-Dine baryogenesis,
heavy DM with mass of $O(1)$ PeV,
and SUSY theories.

\subsection*{Free-streaming velocity of DM}
Since DM is relativistic after the time of DM decoupling in the low reheating temperature scenario,
it might have a cosmologically relevant free-streaming velocity.
If interactions between DM and the thermal plasma are negligible,
the present-day free-streaming velocity of DM
is calculated as
\beq
 v_0 &\simeq& \frac{ \left. \Eth \right\vert_{T = \TR}}{\mDM} \frac{T_0}{\TR} \lmk \frac{g_{*s} (T_0)}{g_{*s} ( \TR )} \rmk^{1/3}, \\
 &\simeq& 8.7 \times 10^{-9} \lmk \frac{\mDM}{1.5\text{ TeV}} \rmk
 \lmk \frac{\TR}{100\text{ MeV}} \rmk^{-2} 
 \lmk \frac{g_{*s} (T_0)}{g_{*s} ( \TR )} \rmk^{1/3}, \\
 &\sim& 8.7 \times 10^{-9} \lmk \frac{\mDM}{1.5 \text{ TeV}} \rmk^{-1/3}
 \lmk \frac{g_{*s} (T_0)}{g_{*s} ( \TR )} \rmk^{1/3}, 
\eeq
where $T_0$ ($\simeq 2.3 \times 10^{-4}$ eV) is the temperature at the present time,
and $g_{*s}$ is the effective number of relativistic degrees of freedom for entropy.
Here, we assume that $\mphi \geq \mDM^2/2 \TR$, and use Eq.~(\ref{result mDM}) in the last line.
Although the observation of the Lyman-$\alpha$ forest constrains the free-streaming velocity
as $v_0 \lesssim 2.5 \times 10^{-8}$~\cite{Viel:2013fqw} (see Ref.~\cite{Steffen:2006hw} for review),
we find that the above result satisfies this constraint when $\mDM \gtrsim 100$ GeV.
The free-streaming velocity will be further constrained 
by future observations of the redshifted 21-cm line 
because
the erasure of small-scale structure results in delaying star formation and thus 
delaying the buildup of UV and
X-ray backgrounds, which affects the 21-cm radiation signal produced by neutral hydrogen.
It is expected that future observations of the redshifted 21-cm line would improve the upper bound
to $v_0 \lesssim 2 \times 10^{-9}$~\cite{Sitwell:2013fpa}.
The low reheating temperature scenario
with $\mDM \lesssim 100$ TeV would be tested by future observations of the redshifted 21-cm line.
In many cases, however, we have to take into account
interactions between DM and the thermal plasma
and the constraint from free-streaming velocity is absent~\cite{Hisano:2000dz, Arcadi:2011ev, Ibe:2012hr}.

\subsection*{Affleck-Dine mechanism}
Since the reheating temperature is very low,
mechanisms to account for the baryon asymmetry of the Universe
are limited.
One well-motivated mechanism is the Affleck-Dine mechanism,
which is naturally realized in SUSY theories~\cite{AD, DRT}.
Note that
the Affleck-Dine mechanism predicts non-zero baryonic isocurvature fluctuation
when Hubble-induced A terms are absent
during inflation.
This is indeed realized if one considers the models of D-term inflation~\cite{Enqvist:1998pf, Enqvist:1999hv, Kawasaki:2001in},
or if the field which has a non-zero F-term during inflation is charged under some symmetry
and its vacuum expectation value is less than the Planck scale during inflation~\cite{Kasuya:2008xp}.
Since observations of cosmic microwave background have shown that
the density perturbations are predominantly adiabatic~\cite{Hinshaw:2012aka,Ade:2013uln},
the isocurvature perturbation is tightly constrained.
The $Planck$ collaboration puts an upper bound as~\cite{Ade:2013zuv}
\beq
 \left\vert S_{{\rm b} \gamma} \right\vert \lesssim \frac{\Omega_{\rm DM}}{\Omega_{\rm b}} \lmk 0.039 \times 2.2 \times 10^{-9} \rmk^{1/2} 
 \simeq 5.0 \times 10^{-5},
\eeq
where
$S_{{\rm b} \gamma}$ is the baryonic isocurvature fluctuation
and $\Omega_{\rm b}$ denotes the density parameter
of the baryon.
This upper bound then gives a constraint as~\cite{Kasuya:2008xp}
\beq
\TR \lesssim 1.1 \times 10^{-17} 
\frac{1}{n^2} \frac{\Mpl^2}{m_{3/2}} \lmk \frac{\mDM}{H_{\rm inf}} \rmk^{\frac{2n-6}{n-2}}
\Theta,
\eeq
where
$m_{3/2}$ is a gravitino mass, $H_{\rm inf}$ is the Hubble parameter
during inflation, and $\Theta$ is an $O(1)$ factor.
Here, we assume that the Affleck-Dine field $\Phi$ has a mass of the order of $\mDM$ and is stabilized via a superpotential term $\propto \Phi^n~(n\geq 4)$.
Using Eq.~(\ref{result mDM}), we obtain the following constraints:
\beq
\mDM \lmk \frac{m_{3/2}}{\mDM} \rmk^{3/2} \lesssim 10^{14} \ \GeV
\lmk \frac{H_{\rm inf}}{10^{12} \ \GeV} \rmk^{-3/2} \Theta^{3/2},
\eeq
for $n=4$, and
\beq
\mDM \lmk \frac{m_{3/2}}{\mDM} \rmk^{6} \lesssim 2 \times 10^{19} \ \GeV
\lmk \frac{H_{\rm inf}}{10^{12} \ \GeV} \rmk^{-9} \Theta^{6},
\eeq
for $n=6$.
While the Affleck-Dine baryogenesis with $n=6$ puts the severe upper bound on the energy scale of inflation $H_{\rm inf}$,
we can easily avoid this constraint for the case of $n=4$.

\subsection*{PeV DM}
It is worth noting that we can account for the abundance of DM
even in the case that
DM is a WIMP with a mass larger than the unitarity bound of a few hundred TeV~\cite{Griest:1989wd}.
The recent observation of high-energy cosmic-ray neutrinos by the IceCube experiment
may indicate that DM is a long-lived particle
with a mass of $O(1)$ PeV~\cite{Aartsen:2013jdh, Feldstein:2013kka, Esmaili:2013gha, Bai:2013nga}.
The above scenario for non-thermal production of DM can also account for the abundance of even such a heavy DM.

\subsection*{SUSY theories}
SUSY models often have difficulties in obtaining the correct DM abundance.
For example, in the constrained minimal SUSY standard model,
the LSP is bino-like in most of the parameter space, which leads to overclosure by thermally produced binos.
Although this situation can be remedied by co-annihilation~\cite{Griest:1990kh,Gondolo:1990dk} with the stau~\cite{Ellis:1998kh}, fine-tunings are required.
In the low reheating temperature scenario, the bino-like LSP can be consistent with the observed DM density without fine-tunings in the mass spectrum.
For SUSY particles with masses of $O(1)$ TeV, the elestic scattering cross section between the bino-like LSP and nucleon is as large as $10^{-46}-10^{-45} \ {\rm cm}^2$, which is detectable in future direct detection experiments of DM such as XENON1T~\cite{Aprile:2012zx}.

%%%%%%%%%%%%%%%%%%%%%%%%%%%%%%%%%%%%%%%%%%%%%%%%%%%%%%%%%%
\section{\label{conclusion}conclusions}
%%%%%%%%%%%%%%%%%%%%%%%%%%%%%%%%%%%%%%%%%%%%%%%%%%%%%%%%%%

We have considered WIMP DM
in a scenario with low reheating temperature.
Although there are several mechanisms to produce DM in this scenario
including thermal production and production in a shower from the decay of the inflaton,
the DM production by inelastic scatterings
between high-energy particles and the thermal plasma
gives the dominant contribution when the mass of the inflaton is sufficiently large.
We have found that the abundance of DM
depends mainly on the mass scale of the DM sector and the temperature of reheating,
but not on the mass of the inflaton as long as the mass is sufficiently large.
We have also found that the observed DM abundance can be accounted for
when DM is heavier than $O(100)$ GeV.
This conclusion is highly independent of the branching ratio of the inflaton
as long as it decays into Standard Model particles
since high-energy particles split into a lot of particles throughout inelastic scattering processes
and the information of the initial condition is lost.

The above scenario is related to some cosmological topics.
For example, the recent observation of high-energy cosmic-ray neutrinos by the IceCube experiment
may indicate that DM is a long-lived particle with a mass of $O(1)$ PeV.
The above scenario for non-thermal production of DM can also account for the abundance of even such heavy DM.
In addition,
since DM is produced non-thermally after the time of DM decoupling,
it might have a cosmologically relevant free-streaming velocity.
If interactions between DM and the thermal plasma are irrelevant after the DM production,
the present-day free-streaming velocity of DM is non-zero
and would be detected by future observations of the redshifted 21-cm line.

%---------------SECTION------------------%
%
\section*{Acknowledgements}
This work is supported by
Grant-in-Aid for Scientific research 
from the Ministry of Education, Science, Sports, and Culture (MEXT), Japan, No. 25400248 (M.K.), No. 21111006 (M.K.); 
the World Premier International Research Center Initiative (WPI Initiative), MEXT, Japan (K.H., M.K. and M.Y.);
the Program for Leading Graduate Schools, MEXT, Japan (M.Y.);
and JSPS Research Fellowships for Young Scientists (K.H., K.M. and M.Y.).
%
%---------------SECTION------------------%

%%%%%%%%%%%%%%%%%%%%%%%%%%%%%%%%%%%%%%%%%%%%%%%%%%%%%%%%%%
\appendix*
\section{\label{appendix}Boltzmann equation describing splittings and its stable solution}
%%%%%%%%%%%%%%%%%%%%%%%%%%%%%%%%%%%%%%%%%%%%%%%%%%%%%%%%%%

In this Appendix, we solve the Boltzmann equation describing splitting processes
and justify the estimation of the number density of DM in Sec.~\ref{inela}.
Let us consider a system that consists of radiation, the inflaton, and DM.
In splitting processes, perpendicular momenta of daughter particles are negligible.
Therefore, it is convenient to consider a momentum distribution function $\tilde{f}(p,t)$ reduced to one dimension,
such that the number density is given by $n(t) = \int \dd p \tilde{f}(p,t)$.

The Boltzmann equation which controls splitting processes is written as
\beq
\label{boltzmann}
 \frac{\del}{\del t} \tilde{f}_{\text{SM}} \lmk p, t \rmk - 3H p \frac{\del}{\del p} \tilde{f}_{\text{SM}} \lmk p, t \rmk
 = \frac{ \dd \Gphi}{\dd p} n_\phi(t) + \lmk \text{collision term} \rmk,
\eeq
where $\tilde{f}_{\rm SM}$ is the momentum distribution of the radiation, and
\beq
 n_\phi(t) &=& n_\phi (0) \lmk \frac{a(0)}{a(t)} \rmk^3 e^{-\Gphi t}, \\
 \frac{ \dd \Gphi}{\dd p} &=& 2 \Gphi \delta \lmk p - \mphi / 2 \rmk,
\eeq
are the number density of the inflaton and its decay rate, respectively.
The collision term is given as
\beq
 \lmk \text{collision term} \rmk
 &=& - \int_0^{p/2} \dd k \frac{\dd \Gamma_{\rm split}}{\dd k} (k) \tilde{f}_{\rm SM} \lmk p, t \rmk \nonumber\\
 &+& \int_{2p}^{\mphi/2} \dd k \frac{\dd \Gamma_{\rm split}}{\dd p} (p) \tilde{f}_{\rm SM} \lmk k, t \rmk \nonumber\\
 &+& \int_0^p \dd k \frac{\dd \Gamma_{\rm split}}{\dd k} (k) \tilde{f}_{\rm SM} \lmk p+k, t \rmk.
\eeq
(see Fig.~\ref{fig2}), where the rate of the splitting process is given by Eqs.~(\ref{inela rate}) and (\ref{del t 2}).
Here we write it as
\beq
 \frac{\dd \Gamma_{\rm split}}{\dd k} (k) = - \frac{1}{2} A k^{-3/2}, \\
 A \equiv k^{1/2} \Gamma_{\rm split}(k) = \text{(const.)}.
\eeq
In the above collision terms, we neglect the back reaction coming from the DM sector
because the reaction rate of DM production is much smaller than that of splitting into the radiation itself.

When there is a hierarchy among time scales,
\beq
 \Gphi \lesssim H \ll \Gsplit,
\eeq
which is the case of our interest in this paper, 
the Boltzmann equation (\ref{boltzmann}) can be solved in the following way.
Since the time scale of the Hubble expansion is much longer than that of the splitting process,
the second term of the left-hand side of Eq.~(\ref{boltzmann}) is negligible.
Further, since Standard Model particles are continuously supplied by the decay of the inflaton
and immediately participate in splitting processes, $\tilde{f}_{\rm SM}$ becomes a constant in time
(up to a slow variation due to the red shift of the source term) as long as the source term is present,
that is, $\Gphi \lesssim t^{-1}\sim H$.
Then, the Boltzmann equation is reduced to the following equation:
\beq
 &&0 =
 - \half A \lkk \int_0^{p/2} \frac{\dd k}{k^{3/2}} \tilde{f}_{\rm SM} (p) \right. 
 - \left. \int_{2p}^{\mphi /2} \frac{\dd k}{p^{3/2}} \tilde{f}_{\rm SM} (k) 
 - \int_0^{p} \frac{\dd k}{k^{3/2}} \tilde{f}_{\rm SM} (p+k) \rkk \quad \text{for } p < \mphi /4, \nonumber \\
 && 2 n_\phi \Gphi \delta \lmk p - \mphi / 2 \rmk = 
 \half A \lkk \int_0^{p/2} \frac{\dd k}{k^{3/2}} \tilde{f}_{\rm SM} (p) \right.
 - \left. \int_0^{\mphi / 2 - p} \frac{\dd k}{k^{3/2}} \tilde{f}_{\rm SM} (p+k) \rkk \quad \text{for } p > \mphi /4. \nonumber \\ 
 \label{Boltzmann2}
\eeq

Let us focus on the momentum distribution for $p \ll \mphi$ and make an ansatz $\tilde{f}_{\rm SM} (p) \propto p^{-n}$.
The equation is then reduced to
\beq
 0 = - \int_0^{p/2} \frac{\dd k}{k^{3/2}} p^{-n}
 + \int_{2p}^{\mphi /2} \frac{\dd k}{p^{3/2}} k^{-n} 
 + \int_0^{p} \frac{\dd k}{k^{3/2}} (p+k)^{-n}.
\eeq
This can be rewritten as
\beq
 \frac{1}{-n+1} \lmk \lmk \frac{\mphi}{2 p} \rmk^{-n+1} - 2^{-n+1} \rmk + 2 \sqrt{2} + i B_{-1} \lmk - \half, 1-n \rmk = 0, \\
 2 \sqrt{2} + i B_{-1} \lmk - \half, 1-n \rmk 
 = \int_0^1 \frac{\dd x}{x^{3/2}} \lmk 1+x \rmk^{-n} - \int_0^{1/2} \frac{\dd x}{x^{3/2}},
\eeq
where $B$ is an incomplete beta function and $i$ is the imaginary unit.
Although the integrals in the second line have infrared divergence, the sum of them is finite.
This relation implies $n =3/2$ for $p/ \mphi \ll 1$, that is,
\beq
 \tilde{f}_{\rm SM} (p) \simeq \frac{n_\phi \Gphi}{A} \mphi p^{-3/2},
\eeq
where we include the coefficient $n_\phi \Gphi / A$ since the source term is proportional to this factor (see Eq.~(\ref{Boltzmann2})).
We also include a factor of $\mphi$ by dimensional analysis.

To verify this result,
let us confirm
the conservation of energy in the following way.
The distribution function $\tilde{f}_{\rm SM}(p)$ represents the distribution of particles
that is produced from the source and
evolves through inelastic scattering during the time of $O(1/\Gamma_{\rm split})$.
Thus what we should calculate in order to confirm the conservation of energy
is the time integral of the source term
and the momentum integral of the distribution function:
\beq
 \int_{t_0}^{t_0 + 1/ \Gamma_{\rm split}} \dd t \ m_\phi n_\phi \Gamma_\phi
 \leftrightarrow
 \int_0^{\mphi/2} \dd p \ p \tilde{f}_{\rm SM} (p),
\eeq
where $t_0 (< 1/\Gphi)$ is an arbitrary time.
The left-hand side is the total energy produced by inflaton decay during $t_0 < t < t_0 + 1/ \Gamma_{\rm split}$,
and the right-hand side is the total energy of radiation 
that evolves during the time of $O(1/\Gamma_{\rm split})$.
These integrals are both calculated as
\beq
 \frac{\mphi n_\phi \Gphi}{\Gamma_{\rm split}},
\eeq
and this result indicates the conservation of energy.
However, we should note that
this stable solution violates the conservation of energy for longer time span ($\gg 1/ \Gamma_{\rm split}$),
since the energy flows into the thermal plasma after the time of $O(1/\Gamma_{\rm split})$.

We now consider the DM production process.
DM is produced from high-energy particles with energy larger than $\Eth$, 
and the rate of the production process is given as $\GDM$.
Thus, the number density of DM is calculated as
\beq
 n_{\rm DM}(t) \sim  \int_{\Eth}^{\mphi/2} \dd p \tilde{f}_{\rm SM} (p) \GDM t
 \sim n_\phi \Gphi t \frac{ \GDM \mphi}{A \Eth^{1/2}}
 \sim n_\phi \Gphi t \frac{\GDM}{\Gamma_{\rm split} (\Eth)} \frac{\mphi}{\Eth}.
\eeq

Let us assume that the DM production becomes inefficient at a time $t_{\rm e}$,
which is given by kinematics $\left(\mphi T\left(t_{\rm e}\right)\sim \mDM^2\right)$
or the disappearance of the source ($t_{\rm e}\sim \Gphi^{-1}$).
Since the ratio of the production rate of DM to the total reaction rate is given as
%%
%\Red{
\beq
 \frac{\Gamma_{\rm DM}}{\Gamma_{\rm split}} \sim \frac{\alpha_{\text{DM}}^2 T^3}{\mDM^2} \frac{\sqrt{\Eth}}{\alpha^2 T \sqrt{T}},
\eeq
%}
%%
we conclude that the energy density of DM is given by
%%
%\Red{
\beq
 \frac{\rho_{\rm DM}}{s}
 \sim \mDM \frac{\alpha_{\text{DM}}^2 T^3}{\mDM^2} \frac{\sqrt{\Eth}}{\alpha^2 T \sqrt{T}} \frac{\mphi}{\Eth} \Gphi t_{\rm e} \times \frac{n_\phi}{s} 
\eeq
%}
%%
where we divide the energy density by the entropy density, $s$.
The ratio $n_\phi/s$ should be estimated at the reheating.
This result justifies
Eq.~(\ref{entropy ratio}).

%%%%%%%%%%%%%%%%%%%%%%%%%%%%%%%%%%%%

%%%%%%%%%%%%%%%%%%%%%%%%%%%%%%%%%%%%

\end{document}